%% file: main.tex
\documentclass[twoside,final,1p,times]{elsarticle}

\usepackage{amsmath}
\usepackage{amssymb}
\usepackage{booktabs}
\usepackage{caption}
\usepackage{color}
\usepackage{comment}
\usepackage{fancyhdr}
\usepackage{graphicx}
\usepackage{multirow}
\usepackage{pgfplots}
\usepackage{textcomp}
\usepackage{tikz}
\usepackage{url}
\usepackage[table]{xcolor}
\usetikzlibrary{shadows, calc, positioning}
\pgfplotsset{compat=1.18}
\usepackage{adjustbox}

\journal{Data Intelligence}

\newcommand\shorttitle{\textcolor[RGB]{20,180,190}{Data Intelligence}}
\biboptions{numbers,sort&compress}

\begin{document}

\begin{frontmatter}
\title{M-CIF: Multi-Scale Alignment for CIF-Based Non-Autoregressive ASR}

\author[addr1]{Ruixiang Mao}
\author[addr1]{Xiangnan Ma}
\author[addr1]{Qing Yang}
\author[addr1]{Ziming Zhu}
\author[addr1]{Yucheng Qiao}
\author[addr1]{Yuan Ge}
\author[addr1,addr2]{\mbox{Tong Xiao}\corref{cor1}}
\author[addr3]{Shengxiang Gao}
\author[addr3]{Zhengtao Yu}
\author[addr1,addr2]{Jingbo Zhu}

\address[addr1]{School of Computer Science and Engineering, Northeastern University, Shenyang, China}
\address[addr2]{NiuTrans Research, Shenyang, China}
\address[addr3]{Kunming University of Science and Technology, China}
\cortext[cor1]{Corresponding author. Email: xiaotong@mail.neu.edu.cn}

\begin{abstract}
The Continuous Integrate-and-Fire (CIF) mechanism provides effective alignment for non-\hspace{0pt}autoregressive (NAR) speech recognition. This mechanism creates a smooth mapping from acoustic features to target tokens, achieving competitive performance on Mandarin. However, its stability degrades in some languages such as English and French. In this paper, we propose Multi-scale CIF (M-CIF), which performs multi-level alignment by integrating character and phoneme level supervision compressed into subword representations, thereby enhancing alignment. Experiments show that M-CIF reduces WER compared to the Paraformer baseline, especially on CommonVoice by 4.21 in German and 3.05 in French. To further investigate, we define phonetic confusion errors (PE) and space-related segmentation errors (SE) as evaluation metrics. 
Analysis reveals that the phoneme and character layers are essential for specifically improving PE and SE performance, further confirming the interpretability of our multi-scale alignment method.
\end{abstract}


\end{frontmatter}

\section{Introduction}

The Continuous Integrate-and-Fire (CIF) mechanism provides a soft and monotonic alignment strategy for non-autoregressive (NAR) speech recognition~\cite{dong2020cif,gao2022paraformer,yu2021boundary}.
This strategy works by integrating frame-level acoustic features into token-level representations once the accumulated threshold is reached~\cite{dong2020cif}. 
By enabling temporal compression and length prediction, CIF-based models have demonstrated competitive performance on Mandarin~\cite{li2022recent,yu2021boundary,gao2022paraformer,zheng2024efficient,zhang2024cif,lee2018deterministic}. However, their cumulative activation process becomes unstable on languages such as English and French, which feature multi-syllabic pronunciation structures and word boundaries marked by spaces.

\begin{figure}[t]
      \centering
      \includegraphics[
          width=0.8\textwidth,
          trim=0.4cm 0.5cm 0.7cm 0,
          clip
      ]{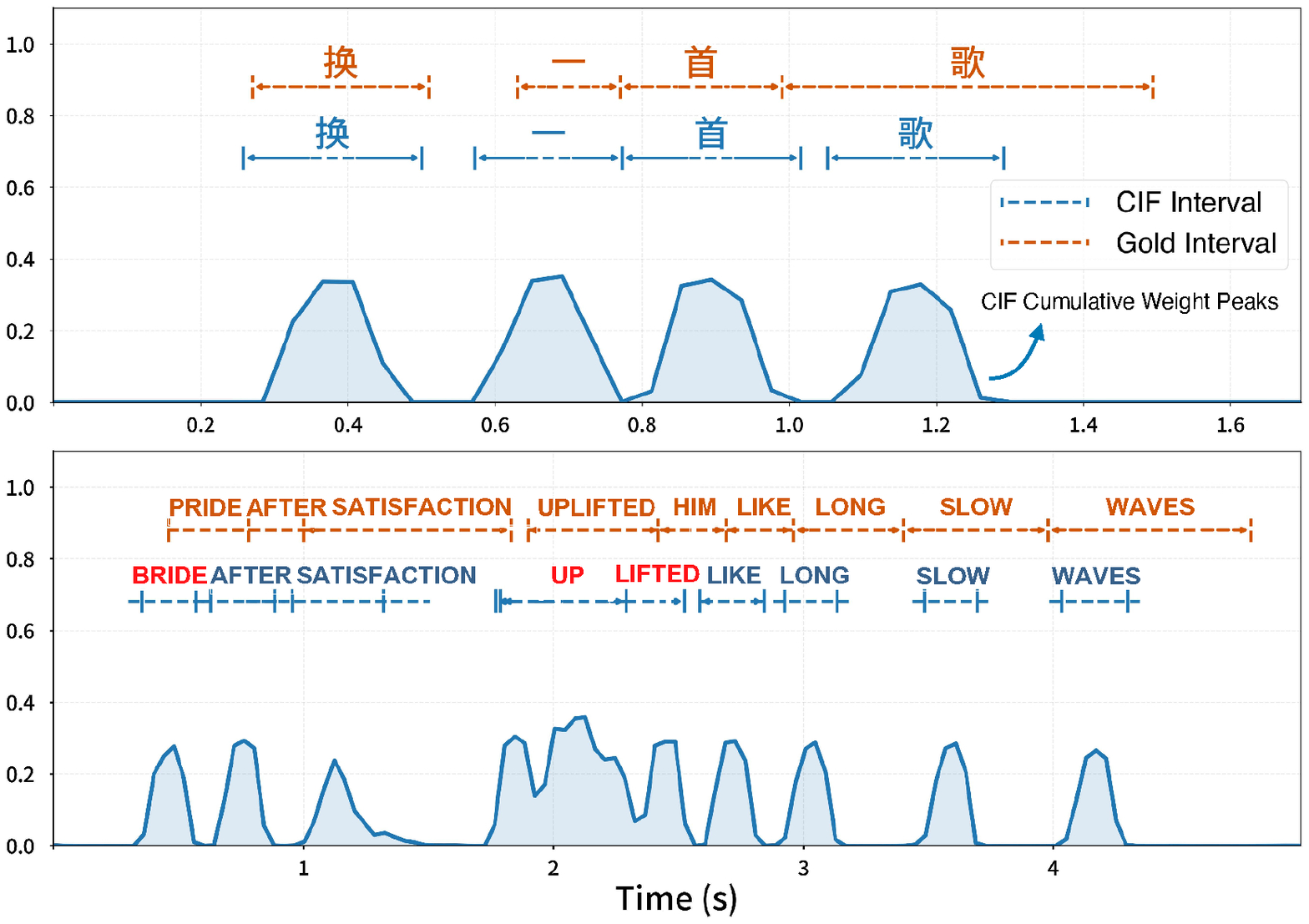}
      \caption{Comparison of feature alignment for CIF and human annotations on a ZH--EN case. Blue and orange
  spans show CIF activations and human references; red marks recognition errors; bottom blue peaks denote
  accumulated CIF weights.}
      \label{fig:merge}
\end{figure}

Specifically, most CIF applications operate at a coarse scale, aligning acoustic-text features primarily at the subword level~\cite{gao2022paraformer,zou2024eparaformer}. Activations occur once features cross a threshold, yet words are treated as indivisible units, disregarding their internal syllabic structure. 
In particular, when encountering multi-syllabic subwords, the lack of finer-grained guidance makes it difficult to capture the inherent acoustic information. 
For example, Mandarin, an isolating language \cite{gil2008complex}, uses words like ``Beijing'' that consist of two clearly separable single-syllabic characters ``Bei'' and ``Jing'', making the CIF alignment process straightforward. 
However, English and French, displaying features of synthetic language \cite{reber1969transfer}, whose core words are composed of multiple pronounced units, such as ``unbelievable'', which contains the prefix ``un-'', the root ``believe'', and the suffix ``-able''. 
As shown in Figure~\ref{fig:merge}, compared to Chinese, English suffers from a more pronounced alignment shift and frequent identification errors at multi-syllabic segments, revealing the instability of CIF on multi-syllabic languages.
This observation substantially challenges CIF acoustic--text alignment in multilingual speech recognition.
 
In this work, we propose Multi-scale CIF  (\textit{M-CIF}), a multi-scale hierarchical framework. Our method progressively compresses and aligns fine-grained character-level and phoneme-level features into coherent subword-level representations in a hierarchical manner, enabling more joint integration across scales. Furthermore, scale-matched CTC losses are applied at each level to provide explicit alignment supervision. 
Implemented within Paraformer~\cite{gao2022paraformer}, \textit{M-CIF} delivers an average relative Word Error Rate (WER) reduction of 0.31 on the LibriSpeech English test set, and up to 4.21 and 3.05 on German and French CommonVoice, respectively. 
Meanwhile, to further investigate the impact of phoneme-level and character-level guidance, we define two error types: phonetic confusion errors (\textit{PE}) and space-related segmentation errors (\textit{SE}). The analysis based on these metrics further confirms the interpretability of our multi-scale alignment method.
We further visualize the predicted alignments against human-annotated timestamps, showing that phoneme-level and character-level guidance is essential for stabilizing CIF alignments in multi-syllabic languages.
Our contributions are:
\begin{figure}[t]
    \centering
    \includegraphics[width=\textwidth,keepaspectratio=true, trim=0cm 0cm 0cm 0cm,clip]{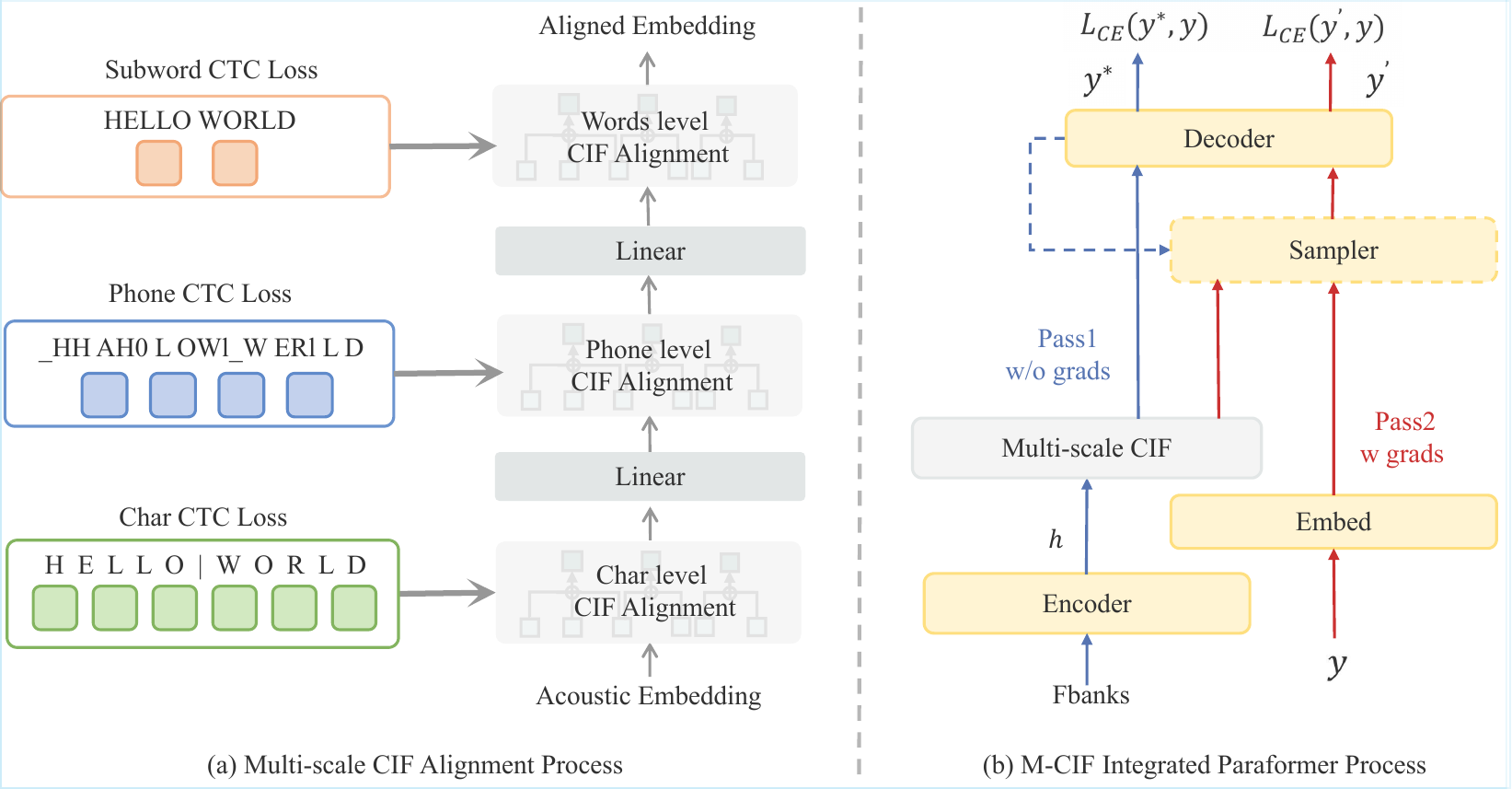}
    \caption{Method overview. (a) Integration of character-level and phoneme-level features into subword-level representations, aligned with scale-matched CTC loss; (b) In Paraformer, M-CIF serves as a finer-grained bridge between the model encoder and decoder.}
    \label{fig:main}
\end{figure}

\begin{itemize}
   \item We propose \textit{M-CIF}, a multi-level alignment framework that augments subword-level features with fine-grained guidance to improve accuracy on languages like English and French.
   \item We define PE and SE metrics to systematically quantify pronunciation confusions and segmentation errors, enhancing multi-scale interpretability.
   \item Our experiments empirically validate \textit{M-CIF}'s performance gains and its effectiveness in mitigating PE and SE errors.
\end{itemize}

\section{Methodology}

In this section, we present a comparative visualization of the CIF firing process in isolating languages and those exhibiting synthetic characteristics languages. Subsequently, after defining PE and SE error types, we introduce the Multi-scale CIF framework to mitigate these challenges.

\subsection{CIF Firing Analysis}
\label{sec:cif-firing-analysis}

As illustrated in Figure \ref{fig:cif}, CIF first predicts frame-level weights $\alpha_{\text{pre}}$ from the acoustic features $H_{\text{Acoustic}}$. It then accumulates these weights until the threshold $\beta$ is reached, emitting the compressed representations $H_{\text{Aligned}}$. This mechanism enables monotonic compression and implicit length modeling, making it widely applicable in both attention-based autoregressive (AR)~\cite{teo2024cif,han2023knowledge} and NAR approaches for ASR~\cite{gao2022paraformer,zou2024paraformer,zheng2024efficient,dong2023cif}.

To investigate differences between languages, we visualize in Figure~\ref{fig:merge} how CIF-predicted weights accumulate to indicate the temporal spans of subwords. Then we compare these predicted spans with manually annotated ground-truth intervals. The visualizations show that CIF activations align  accurately with ground-truth spans in Mandarin, but become unstable in English, which exhibits characteristics of synthetic languages. 
We argue that this instability stems from their multi-syllabic structures and invisible space delimiter structure, which increase the alignment difficulty of CIF and further degrade recognition accuracy. 
Furthermore, the phonetic confusion errors (PE) and space-related segmentation errors (SE) frequently occurr, which is shown in the red-marked regions of Figure~\ref{fig:merge}.

\input{image/fig-2}

To quantify these errors, we compute their rates by normalizing error counts with respect to the number of reference units. 
We use the normalized Levenshtein~\cite{yujian2007normalized} distance as $\text{NLD}~(x,y)=\text{Lev}(x,y)/\max(|x|,|y|)$. PE are counted when the phone distance falls below $\theta_{\text{PE}}$, and SE are counted when the reference and hypothesis show boundary mismatches but their de-spaced strings have a character-level distance below $\theta_{\text{SE}}$.

The PE rate and SE rate are computed as follows:
\begin{equation}
\text{PE Rate} = 
\frac{\text{Count}\{ \text{NLD}(\mathrm{ref}_{\text{phone}}, \mathrm{hyp}_{\text{phone}}) \leq \theta_{\text{PE}} \}}
{\sum{\mathrm{ref}_{\text{phone}}}}
\end{equation}
\begin{equation}
\text{SE Rate} = 
\frac{\text{Count}\{ \text{SE $\cap$ NLD}(\mathrm{ref}_{\text{char}}, \mathrm{hyp}_{\text{char}}) \leq \theta_{\text{SE}} \}}
{\sum{\mathrm{ref}_{\text{boundary}}}}
\end{equation}

\subsection{Multi-scale CIF Strategy}
\label{sec:multi-scale-cif-strategy}
To address the unstable behavior, we propose \textit{M-CIF}, a multi-scale framework that alleviates multi-syllabic errors through multi-level alignment. As shown in Figure \ref{fig:main} (a), it aligns acoustic features at the character, phoneme, and subword levels, with scale-specific CTC objectives providing auxiliary supervision.

\textbf{M-CIF Alignment Strategy}~ Let the encoder output be \( \mathbf{h} = (h_1, h_2, \ldots, h_T) \) and the target transcription be \( \mathbf{Y} = (y_1, y_2, \ldots, y_U) \). In \textit{M-CIF}, the compression is carried out hierarchically through three stages of CIF alignment, operating respectively at the character level ``c'', the phoneme level ``p'', and the subword level ``w''. 
At each stage $s \in \{c, p, w\}$, alignment is obtained by accumulating the weight $\alpha^s$ until a threshold $\beta$ is reached, and then the compressed features as the input for the next stage, formally defined as:
\begin{equation}
\alpha^s = \mathrm{Sigmoid} (\mathrm{Linear}(\mathrm{Conv}(\mathbf{h}^s)))
\end{equation}
\begin{equation}
\mathbf{h}^{s+1} = \mathrm{CIF} (\mathbf{h}^s, \alpha^s)
\end{equation}

Meanwhile, the MAE loss~\cite{willmott2005advantages} is used to constraint the predicted number of emissions to match the ground-truth length $U_s$:
\begin{equation}
\mathcal{L}_{\mathrm{QUA}} = \sum_{s \in \{c,p,w\}} \left| \sum_{t=1}^{T_s} \alpha_t^s - U_s \right|
\end{equation}

In parallel, a multi-scale CTC loss\cite{graves2006connectionist} is applied before each CIF stage, where the weight $W_s$ controls its contribution, thereby providing acoustic supervision at the corresponding granularity. $W_s$ is scheduled across training: supervision begins with stronger emphasis on character alignment, gradually shifts toward phoneme guidance, and ultimately converges on subword constraints in the later stages, calculated by:\begin{equation}
\mathcal{L}_{\mathrm{CTC}} = \sum_{s \in \{c,p,w\}} W_s \cdot \big( -\log P(Y_s \mid h_s) \big)
\end{equation}

Finally, the overall training criterion of \textit{M-CIF} integrates both objectives, combining the multi-scale quantity constraint with the multi-scale CTC regularization:
\begin{equation}
\mathcal{L}_{\mathrm{M-CIF}} = \mathcal{L}_{\mathrm{QUA}} + \mathcal{L}_{\mathrm{CTC}}
\end{equation}

\textit{Char level CIF}~ In languages such as English and French, character-level CIF decomposes words into characters with ``\texttt{|}'' marking boundaries, while in isolating languages like Chinese it operates on processed pinyin. The resulting lengths define the activation targets, with CTC loss applied to stabilize alignment.

\textit{Phoneme level CIF}~ At the phoneme level, we convert text into phonemic sequences using trained G2P models with a Phoneme Error Rate (PER)~\cite{lee2002speaker} of less than 10\%. Building on character-level compressed acoustic features, CIF activations are constrained by phoneme lengths, with phonemes explicitly serving as targets for CTC training.

\textit{Subword level CIF}~ At the subword level, BPE~\cite{kudo2018sentencepiece} tokenization is trained on corpora with a 10k vocabulary, while isolating languages such as Chinese are segmented at the character level. A subword-level CTC constraint is likewise applied before CIF to regularize the compressed acoustic features during training.

\textbf{Model Architecture} ~We implement \textit{M-CIF} on the widely adopted Paraformer~\cite{gao2022paraformer} framework. Paraformer employs a Conformer-based encoder~\cite{gulati2020conformer} and a Transformer-\hspace{0pt}based decoder~\cite{vaswani2017attention}, together with a subword-level CIF module that provides explicit length prediction and enforces monotonic acoustic-to-text alignment. On top of this, a GLM-based sampler, as illustrated in Figure~\ref{fig:main}~(b), generates an initial candidate sequence by sampling from the predicted token distribution, which then serves as the starting point.

\textbf{Cascaded Integration Across Scales}~ 
In previous work, hierarchical and intermediate
CTC tap auxiliary heads at different \emph{depths} of a shared encoder, where every head still operates on a sequence of the same temporal length $T$. 
\textit{M-CIF} is different: it performs the multi-scale alignment \emph{inside the CIF integration path}. 
Through the cascade ${h}^{s+1}=\mathrm{CIF}(h^{s},\alpha^{s})$ for $s\in\{c,p,w\}$, each
scale physically resamples the sequence to the length of the next, coarser unit, so the character-compressed representation feeds the phoneme stage and the phoneme-compressed representation feeds the subword stage. 
Thus, the hierarchy is a sequence of explicit length-compressing integration steps rather than parallel heads
on a fixed-length feature map, and the scale-matched CTC losses merely anchor each stage to its linguistic unit. 
This cascaded compression is what progressively distills character- and phoneme-level acoustic evidence into the subword representation that drives length prediction and decoding, and is the key distinction from depth-wise hierarchical CTC.

\section{Experiments}
\label{sec:experiments}

\subsection{Data and Settings}
\label{sec:data-and-settings}
\textbf{Datasets}~~For a comprehensive multilingual assessment of \textit{M-CIF}, we conduct experiments across multiple datasets. These include 960 hours of English data from LibriSpeech \cite{panayotov2015librispeech}, as well as 950 hours of German and 830 hours of French from the CommonVoice corpus \cite{ardila2019common}. Additionally, we utilize a combined total of 1,150 hours of Chinese speech from AISHELL-1 \cite{bu2017aishell} and AISHELL-2 \cite{reber1969transfer}. All models extract 80-dimensional filter banks as acoustic input features.

\textbf{Baseline}~~We select Paraformer and its variant E-Paraformer~\cite{zou2024eparaformer} as our baselines, and integrate the proposed \textit{M-CIF} framework into Paraformer. Compared to basic Paraformer, which employs the base CIF structure, E-Paraformer further introduces the Parallel Integrate-and-Fire (PIF) mechanism, replacing CIF's recursive alignment with a parallel procedure that computes a global attention matrix in one shot. For all models, we employ a 12-layer Conformer encoder and a 12-layer Transformer decoder, each with a hidden size of 256.

\textbf{Training}~~During the training stage, we employ a hyperparameter scheduling strategy tailored for the multi-scale architecture. CTC losses at different CIF levels are weighted with a scheduled emphasis across stages, while a learning-rate annealing scheme is applied: after 90 epochs, the learning rate is reset to $6.448 \times 10^{-5}$ and subsequently decayed to promote stable and efficient convergence. To stabilize training on languages like Chinese, where token lengths across structural levels are relatively close, we adopt a three-stage curriculum~\cite{zhang2025soundwave}. 
Stage I optimizes the character-level CTC and quantity losses. Stage II additionally introduces the phoneme-level CTC and quantity losses. Stage III further adds the subword-level CTC and quantity losses together with the final decoder cross-entropy loss. After 90 epochs, the learning rate is reset to $6.448\times10^{-5}$ and is subsequently decayed. This curriculum is used empirically to stabilize the optimization of the cascaded alignment stages.

For our experiments, all implementations are based on the open-source FunASR~\cite{gao2023funasr} toolkit\footnote{The tool is available at https://github.com/modelscope/FunASR}. The acoustic features are augmented using SpecAugment~\cite{park2019specaugment}, and training is conducted for 150 epochs on EN, DE and FR dataset and 50 epochs on isolating language dataset with eight NVIDIA 3090 GPUs. 

\subsection{Overall Performance}
\label{sec:overall-performance}

Integrating multi-scale CIF into Paraformer yields consistent improvements across multi-syllabic languages such as English and French. 
As shown in Table~\ref{tab:result}, relative WER reductions of 0.31 are observed on average for the LibriSpeech test sets, together with reductions of 3.05 on the French and 4.21 on the German CommonVoice test sets. 
\input{table/table3}
On Chinese corpora, the WER difference is only around 0.18 relative to the baseline, suggesting that single-scale CIF already works effectively for single-syllable languages where alignment boundaries are naturally stable. 
Notably, even the single subword CTC variant (three-scale CIF with a single subword-level CTC) already surpasses the standard baseline, confirming the validity of multi-scale CIF itself. 
The total parameter count increases from 60.11M to 65.39M, an increase of 5.28M (8.8\%). The auxiliary CTC projection heads are removed after training, whereas the cascaded CIF modules remain in the inference path. Accordingly, the measured RTF increases from 0.017 to 0.022, corresponding to an absolute increase of 0.005 and a relative increase of 29.4\%. M-CIF therefore introduces a measurable inference cost in exchange for the WER improvements on English, French, and German.

M-CIF thus improves accuracy while remaining roughly an order of magnitude faster than autoregressive decoding.

It is worth clarifying that we study alignment within the non-autoregressive (NAR) family rather than comparing against autoregressive (AR) models such as Conformer. 
AR models decode sequentially and incur high latency, whereas NAR models predict in parallel and are about 10$\times$ faster, typically at some cost in accuracy; on Mandarin, Paraformer already matches most AR baselines. 
A direct AR comparison would therefore conflate two distinct speed–accuracy trade-offs, so we instead focus on improving multilingual performance within the NAR setting. 
The gains of M-CIF stem primarily from the hierarchical CIF design rather than from auxiliary supervision alone. 
Applying hierarchical CTC directly to a single-level feature induces instability from conflicting gradients.
M-CIF avoids this by first compressing acoustic features hierarchically, so that each CTC loss is aligned with its corresponding linguistic level.

\input{table/PE_SE}

\section{Analysis}
\label{sec:analysis}

\subsection{Ablation Study}
\label{sec:ablation-study}

We perform ablation experiments by removing the phoneme and character level alignments while keeping other settings unchanged, with a configuration retaining only the single CTC.
Our ablation results in Table~\ref{tab:result} reveal that removing either the character layer or the phoneme layer consistently increases WER in English, French, and German. 
Meanwhile, keeping the three-layer CIF with a single-layer CTC also increases WER across multiple languages, but the performance drop is smaller compared to the other settings. 
These results indicate that each component contributes complementarily to the overall performance, while the three-layer CIF serves as the primary mechanism for alignment.
Based on this, the multi-scale CIF framework performs hierarchical alignment, where character and phoneme level supervision is progressively compressed into coherent subword-level representations. 
This layered design sharpens alignment by internalizing fine-grained phonological and boundary information, ultimately improving the final representations and reducing WER.

\subsection{PE and SE Metrics Analysis}
\label{sec:pe-se-analysis}
We conduct a detailed comparative analysis based on the ablation results, focusing specifically on PE and SE. As summarized in Table~\ref{tab:PE-SE}, the Paraformer baseline shows that both error types occur frequently in multi-syllabic languages, indicating that single-level CIF produces unstable and imprecise alignments with abundant PE and SE errors. By contrast, \textit{M-CIF} framework substantially reduces both types of errors, demonstrating its effectiveness in addressing multi-syllabic confusions and boundary errors.
However, compared to French and German, English exhibits marginal improvements in WER, PE, and SE metrics. We attribute this to the fact that modern English retains fewer synthetic language characteristics~\cite{baugh1993history} than French and German, namely multi-syllabic structures and space-delimited word boundaries. The diminished presence of these features is likely one of the reasons for the limited performance gains.

\textbf{PE Metrics}~~As shown in Table \ref{tab:PE-SE}, on the English clean set and the German and French test sets, removing the phoneme layer leads to a sharper rise in PE rates than removing the character layer. This underscores the stronger corrective role of phoneme-level guidance in mitigating phonetic confusions. Furthermore, in German and French, the setting with only phoneme and subword layers achieves the lowest PE rates, reflecting that preserving phoneme alignment provides the most effective reduction of such errors.

\textbf{SE Metrics}~~Table \ref{tab:PE-SE} shows that the full multi-scale CIF yields the lowest SE rates, strongly demonstrating that reliable word-boundary segmentation in languages like German requires the combined effect of character guidance. Furthermore, SE rates rise markedly more when the character layer is removed than when the phoneme layer is ablated. This confirms that fine-grained orthographic supervision exerts a stronger corrective influence on segmentation errors.

  \begin{figure}[t]
      \centering
      \includegraphics[
          width=0.9\textwidth,
          trim=0.35cm 4.85cm 1cm 4cm,
          clip
      ]{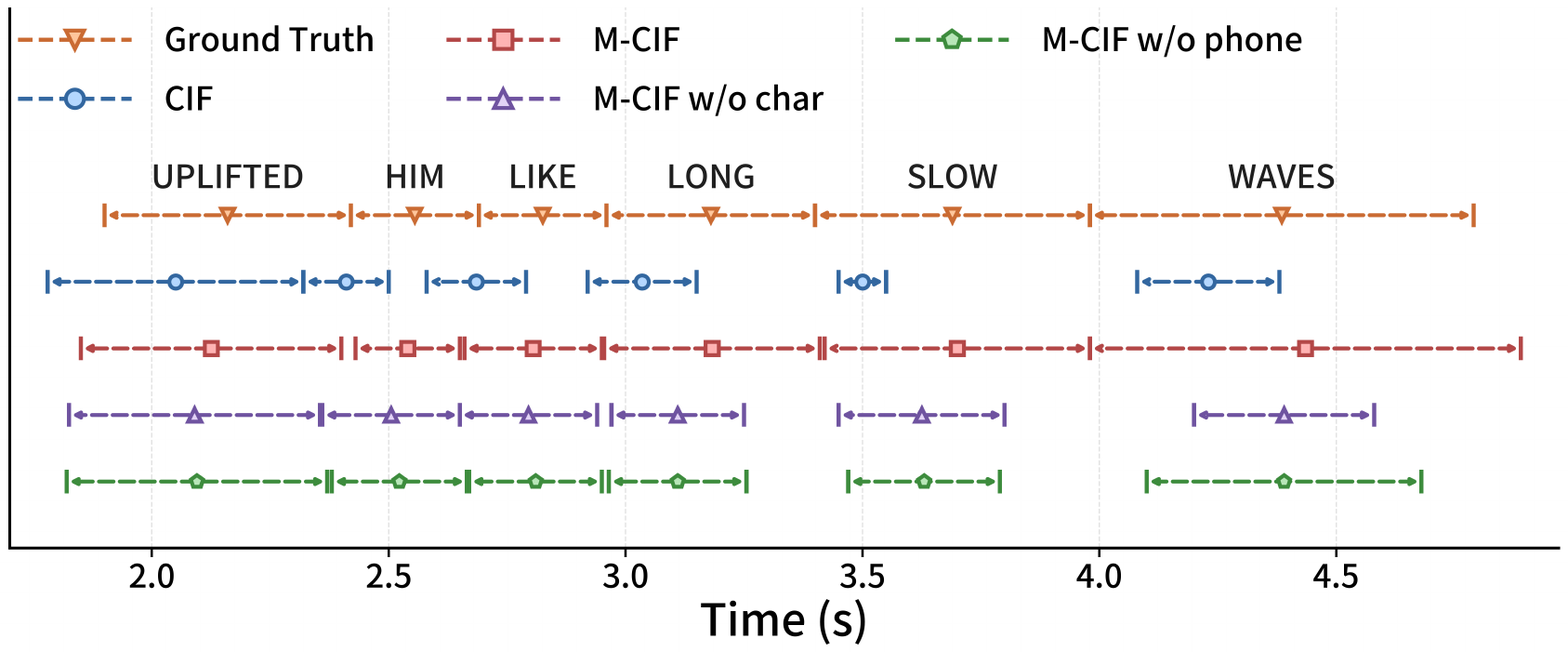}
      \caption{Comparison of text--timestamp alignments between different M-CIF settings and the human
  annotations.}
      \label{fig:timestamp}
  \end{figure}

\subsection{CIF Text-timestamp Alignment Analysis}
\label{sec:timestamp-analysis}
To further investigate \textit{M-CIF}'s effectiveness in improving alignment for languages like English, we present a comparative visualization against human-annotated timestamps. 
As shown in Figure \ref{fig:timestamp}, the complete \textit{M-CIF} configuration aligns most closely with the ground-truth timestamps. This demonstrates that the multi-level design markedly improves alignment fidelity in languages such as English. Meanwhile, the ablated variants that remove either the phoneme layer or the character layer achieve better alignment than the original CIF but still lag behind the full configuration. These results indicate that incorporating phoneme-level and character-level guidance is essential for stabilizing CIF alignments in multi-syllabic  languages.

\section{Related Work}
\label{sec:related-work}

\textbf{Continuous Integrate-and-Fire}~
The Continuous Integrate-and-Fire (CIF) mechanism~\cite{dong2020cif} provides a soft, monotonic acoustic-to-text alignment by accumulating frame-level weights until a threshold is reached and then firing a token-level representation. It has
been widely adopted in NAR systems such as Paraformer~\cite{gao2022paraformer}
and its parallel variant E-Paraformer~\cite{zou2024eparaformer}, and refined by
boundary and context aware training~\cite{yu2021boundary}, transducer-style
integration (CIF-T)~\cite{zhang2024cif}, hierarchical knowledge
distillation~\cite{han2023knowledge}, pre-training for spoken language
understanding (CIF-PT)~\cite{dong2023cif}, and streaming
recognition~\cite{teo2024cif}. A common property of these systems is that the CIF
module operates at a \emph{single}, coarse (subword level)
granularity: words are integrated as indivisible units, and no finer-grained
acoustic evidence guides the firing process. \textit{M-CIF} instead makes the
integration path itself multi-scale.

\textbf{Non-autoregressive ASR}~
Beyond CIF, NAR-ASR has been explored through CTC refinement and mask-prediction
approaches such as Mask-CTC~\cite{higuchi2020mask}, Imputer~\cite{chan2020imputer},
Align-Refine~\cite{chi2021alignrefine}, and CTC-alignment-based single-step
decoding (CASS-NAT)~\cite{fan2021cassnat}, complementing earlier iterative refinement NAR sequence models~\cite{lee2018deterministic}. These methods focus on
how to iteratively recover token content given an alignment, whereas CIF-based NAR
focuses on \emph{producing} the alignment and target length. \textit{M-CIF}
follows the CIF-based NAR line and improves the alignment stage itself by making
it hierarchical, which is orthogonal to and could be combined with refinement-based
decoders.

\section{Conclusion}
In this work, we propose \textit{M-CIF}, a multiscale framework for languages like English and French. This method progressively compresses fine-grained character-level and phoneme-level features into subword-level representation with scale-matched CTC supervision. Building on this design, it constructs a multi-scale acoustic feature capture alignment process, thereby enhancing speech recognition performance. Experiments on English, French, and German show consistent WER reductions. We further define and analyze phonetic confusion errors (PE) and space-related segmentation errors (SE). Our analysis reveals that the primary advantage of \textit{M-CIF}'s multi-level alignment, which mitigates the challenges of CIF acoustic--text alignment in multilingual speech recognition.

\section*{Acknowledgements}
This work was supported in part by the National Science Foundation of China (Nos. 62276056 and U24A20334), the Yunnan Fundamental Research Projects (No.202401BC070021), the Yunnan Science and Technology Major Project (No. 202502AD080014), the Liaoning Provincial Science and Technology Plan Project (No. 2026JH40/10100033), the Fundamental Research Funds for the Central Universities (Nos. N25BSS054 and N25BSS094), and the Program of Introducing Talents of Discipline to Universities, Plan 111 (No.B16009).

\end{document}

%% file: image/fig-2.tex
\definecolor{myLightBlue}{RGB}{220,225,226}
\begin{figure}[t]
      \centering
      \resizebox{0.82\textwidth}{!}{%
\begin{tikzpicture}[
    h_node/.style={
        draw, thick,
        rounded corners=3pt,      
        fill=myLightBlue,         
        minimum size=1.8em,
        drop shadow
    },
    node distance=2.5cm and 1cm 
]
    \node[scale=1.2] (label_ha) at (-2.7, 2.5) {$H_{\text{Aligned}}$};
    \node[scale=1.2] (label_hac) at (-2.7, 0.5) {$H_{\text{Acoustic}}$};
    \node[scale=1.2] (label_alpha) at (-2.7, -0.2) {$\alpha_{\text{pre}}$};

    \node[h_node] (al1) at (0, 2.5) {};
    \node[h_node] (al2) at (3, 2.5) {};
    \node[h_node] (al3) at (6.8, 2.5) {};

    \node[h_node, below= 1.5cm of al1, label={[below=0.7cm]0.5}] (g1n2) {};
    \node[h_node, left=0.8cm of g1n2, label={[below=0.7cm]0.3}] (g1n1) {};
    \node[h_node, right=0.8cm of g1n2, label={[below=0.7cm]0.3},left color=myLightBlue, right color=white] (g1n3) {};
    
    \node[h_node, below=1.5cm of al2, label={[below=0.7cm]0.6}] (g2n2) {};
    \node[h_node, right=0.8cm of g2n2, label={[below=0.7cm]0.4}] (g2n3) {};

    \node[h_node, below=1.5cm  of al3, xshift=-0.9cm, label={[below=0.7cm]0.6}] (g3n1) {};
    \node[h_node, right=1.2cm of g3n1, label={[below=0.7cm]0.2}] (g3n2) {};

    \node[circle, draw, inner sep=3pt, yshift=-0.05cm] at ($(al1.south)!0.5!(g1n2.north)$) {};
    \node[circle, draw, inner sep=3pt, yshift=-0.05cm] at ($(al2.south)!0.5!(g2n2.north)$) {};
    \node[circle, draw, inner sep=3pt, below=0.65cm of al3] (c1) {};

    \draw[rounded corners,thick] (g1n1.north) -- ([yshift=2.2em]g1n1.north) -- ([yshift=2.2em,xshift=-0.3em]g1n3.north) -- ([xshift=-0.3em]g1n3.north);
    
    \draw[rounded corners,thick] ([xshift=0.3em]g1n3.north) -- ([yshift=2.2em,xshift=0.3em]g1n3.north) -- ([yshift=2.2em]g2n3.north) -- (g2n3.north);
    
    \draw[->,rounded corners,thick] (g1n2.north) -- (al1.south);
    \draw[->,rounded corners,thick] (g2n2.north) -- (al2.south);

    \draw[rounded corners,thick] (g3n1.north) -- ([yshift=2.2em]g3n1.north) -- ([yshift=2.2em]g3n2.north) -- (g3n2.north);
    \draw[->,rounded corners,thick] (c1.south) -- (al3.south);

    \node[scale=0.8] at (-1.12,0.95) {$\times 0.3$};
    \node[scale=0.8] at (-0.38,0.95) {$\times 0.5$};
    \node[scale=0.8] at (1,0.95) {$0.2 \times$};
    \node[scale=0.8] at (1.9,0.95) {$\times 0.1$};
    \node[scale=0.8] at (2.65,0.95) {$0.6 \times$};
    \node[scale=0.8] at (4.1,0.95) {$0.3 \times$};
    \node[scale=0.8] at (5.55,0.95) {$0.6 \times$};
    \node[scale=0.8] at (7.4,0.95) {$0.2 \times$};
    
    
    \node at (0.4,1.7) {$\ge \beta$};
    \node at (3.4,1.7) {$\ge \beta$};
    \node at (7.4,1.7) {$\ge \beta_{\text{tail}}$};
    
\end{tikzpicture}
}
\caption{In the CIF process, the feed-forward network predicts $\alpha_{\text{pre}}$; the threshold $\beta$ is set to 1, with $\beta_{\text{tail}}$ set to 0.45.}
\label{fig:cif}
\end{figure}

%% file: table/table3.tex
\begin{table}[t]
\centering
\small

\caption{WER~(\%) and RTF results of our method, where \textit{w/o Char CIF+CTC} and \textit{w/o Phone CIF+CTC} denote two-scale training without the character or phoneme CIF and CTC. Single subword CTC means a three-layer CIF with a single subword-level CTC loss.
\textit{M-CIF*} denotes the M-CIF method applied in Paraformer. \textit{LS} denotes the setting trained and tested on the LibriSpeech dataset, \textit{CV} denotes the CommonVoice dataset, and \textit{AS2} denotes the AISHELL-2 dataset. All results in the same language were trained under the same settings. The same abbreviations are used throughout the paper.}

\resizebox{\textwidth}{!}{
\setlength{\tabcolsep}{2mm}{
\begin{tabular}{lcccccccc}
\toprule
\multirow{2}{*}{\textbf{Method}} &
\multirow{2}{*}{\textbf{Param.}} &
\multicolumn{3}{c}{\textbf{EN(LS)~$\downarrow$}}  & 
\multirow{2}{*}{\textbf{FR(CV)~$\downarrow$}} &
\multirow{2}{*}{\textbf{DE(CV)~$\downarrow$}} &
\multirow{2}{*}{\textbf{ZH(AS2)~$\downarrow$}} &
\multirow{2}{*}{\textbf{RTF(EN)}} 
\\
\cline{3-5}
~ & ~ & \textbf{\textit{clean}} & \textbf{\textit{other}} & \textbf{\textit{Avg.}} & ~ & ~ & ~ & ~ \\  

\midrule
Paraformer & 60.11 M & 5.67 & 12.04 & 8.86 & 21.80 & 19.48 & \textbf{7.06} & 0.017\\
E-Paraformer & 57.54 M & 8.68 & 18.76 & 13.72 & 30.92 & 27.16 & 15.67 & \textbf{0.007}\\

\hline

\textbf{Our M-CIF*}& 65.39 M & \textbf{5.33} & \textbf{11.76} & \textbf{8.55} & \textbf{18.75} & \textbf{15.27} & 7.24  & 0.022 \\

w/o Char CIF+CTC & 62.75 M & 7.04 & 13.73 & 10.39 ($\uparrow$ 1.84) & 20.75 ($\uparrow$ 2.00) & 16.51 ($\uparrow$ 1.24) & -  & -\\
w/o Phone CIF+CTC & 62.75 M & 6.61 & 12.78 & 9.70 ($\uparrow$ 1.15) & 21.71 ($\uparrow$ 2.96) & 17.07 ($\uparrow$ 1.80) & -  & -\\
single subword CTC & 60.25 M & 6.43 & 12.40 & 9.42 ($\uparrow$ 0.87) & 19.46 ($\uparrow$ 0.71) & 16.24 ($\uparrow$ 0.97) & -  & -\\

\bottomrule
\end{tabular}
}
}

\label{tab:result}
\end{table}

%% file: table/PE_SE.tex
\begin{table}[t]
  \centering
  \small
  \setlength{\tabcolsep}{4pt}

  \caption{Comparison of PE and SE error rates results (values in \textperthousand) for different model
  implementations, with $\theta_{\text{PE}}=0.6$ and $\theta_{\text{SE}}=0.5$. The best performance in each
  group is in \textbf{bold} and the second best one is \underline{underlined}.}

  \begin{adjustbox}{max width=\linewidth}
  \begin{tabular}{lccccc}
  \toprule
  & \multicolumn{3}{c}{\textbf{EN(LS)~$\downarrow$}} & \multirow{2}{*}{\textbf{DE(CV)~$\downarrow$}} &
  \multirow{2}{*}{\textbf{FR(CV)~$\downarrow$}} \\
  \cline{2-4}

  \multirow{-2}{*}{\textbf{Model}} & \textbf{\textit{clean}} & \textbf{\textit{other}} & \textbf{\textit{Avg.}}
  \\
  \midrule
  \multicolumn{6}{c}{\bfseries\itshape Phonetic Confusion Errors (PE)} \\
  \addlinespace[-1pt]
  \midrule

  Base & \underline{29.42} & \underline{41.04} & \underline{35.23} & 74.40 & 58.91 \\
  \textbf{M-CIF*} & \textbf{27.40} & 41.84 & \textbf{34.62} & \underline{68.15} & 58.37 \\
  w/o Char CIF & 31.60 & 43.31 & 37.46 & \textbf{67.34} & \textbf{56.62} \\
  w/o Phone CIF & 31.85 & \textbf{40.95} & 36.40 & 76.07 & \underline{57.26} \\

  \midrule
  \multicolumn{6}{c}{\bfseries\itshape Space-related Segmentation Errors (SE)} \\
  \addlinespace[-1pt]
  \midrule

  Base & \underline{7.37} & \underline{12.53} & \underline{9.95} & 27.14 & \underline{24.36} \\
  \textbf{M-CIF*} & \textbf{7.21} & \textbf{12.02} & \textbf{9.62} & \textbf{21.79} & \textbf{20.54} \\
  w/o Char CIF & 9.51 & 13.89 & 11.70 & 23.44 & 25.23 \\
  w/o Phone CIF & 8.19 & 13.32 & 10.76 & \underline{23.02} & 25.73 \\

  \bottomrule
  \end{tabular}
  \end{adjustbox}

  \label{tab:PE-SE}
  \vspace{-0.2cm}
  \end{table}